\documentclass[reprint,twocolumn]{revtex4-1}
\usepackage{graphicx}%
\usepackage{dcolumn}%
\usepackage{bm}%
\usepackage{amssymb}
\usepackage{epsfig,colordvi}
\usepackage[latin1]{inputenc}
\usepackage{color} 
\usepackage{here}

 \def\be{\begin{equation}}
 \def\ee{\end{equation}}
 \def\bea{\begin{eqnarray}}
 \def\eea{\end{eqnarray}}
 \def\bean{\begin{eqnarray*}}
 \def\eean{\end{eqnarray*}}
 \def\gsim{\mathrel{\rlap{\lower0.2em\hbox{$\sim$}}\raise0.2em\hbox{$>$}}}
 \def\ksim{\mathrel{\rlap{\lower0.2em\hbox{$\sim$}}\raise0.2em\hbox{$<$}}}
 \def\kg{\mathrel{\rlap{\lower0.25em\hbox{$>$}}\raise0.25em\hbox{$<$}}}

 \def\bm#1{\mbox{\boldmath$#1$}}

\newcommand{\AuAu} {$\mbox{$^{197}Au$}+\mbox{$^{197}Au$}$ }

\begin{document}

\title{On the Origin of the Elliptic Flow and its Dependence on the Equation of State in Heavy Ion Reactions at Intermediate Energies}

\author{A. Le F\`evre$^1$, Y. Leifels$^1$, C. Hartnack$^2$  and J. Aichelin$^{2,3}$}
\affiliation{$^1$ GSI Helmholtzzentrum f\"ur Schwerionenforschung GmbH,
  Planckstr. 1, 64291 Darmstadt, Germany} 
\affiliation{$^2$ SUBATECH, IMT Atlantique, Universit\'e de Nantes, IN2P3/CNRS
\\ 4 rue Alfred Kastler, 44307 Nantes cedex 3, France}
\affiliation{$^3$ Frankfurt Institute for Advanced Studies, Ruth Moufang Str. 1
\\ 60438 Frankfurt, Germany}

\date{\today}

\begin{abstract} \noindent
Recently it has been discovered that the elliptic flow, $v_2$, of composite charged
particles emitted at midrapidity in Heavy-Ion collisions at intermediate energies  
shows the strongest sensitivity to the Nuclear Equation of State (EoS) which has been observed up to now within
a microscopic model. This dependence on the nuclear EoS is predicted by Quantum Molecular Dynamics
(QMD) calculations \cite{Fevre:2015fza} which show as well that the absorption or rescattering  of in-plane emitted
particles by the spectator matter is not the main reason for the EoS
dependence of the elliptic flow at mid-rapidity but  different density
gradients (and therefore different forces) in the direction of the impact
parameter (x-direction) as compared to the direction perpendicular to the
reaction plan (y-direction), caused by the presence of the spectator matter. 
The stronger density gradient in y-direction accelerates the particles more
and creates therefore a negative $v_2$. When using a soft momentum dependent EoS,
the QMD calculations reproduce the experimental results.
\end{abstract}


\maketitle

\section{Introduction}

The elliptic flow at midrapidity, originally called out-of-plane emission or
squeeze-out, has attracted a lot of attention during 
the last years.  It has been predicted in hydrodynamical simulations of heavy
ion reactions \cite{Stoecker:1981pg,Buchwald:1984cp,Stoecker:1986ci} and has
later  been found experimentally by the Plastic Ball collaboration
\cite{Gutbrod:1989gh}. 

\begin{figure}
\begin{center}
\includegraphics[width=\columnwidth]{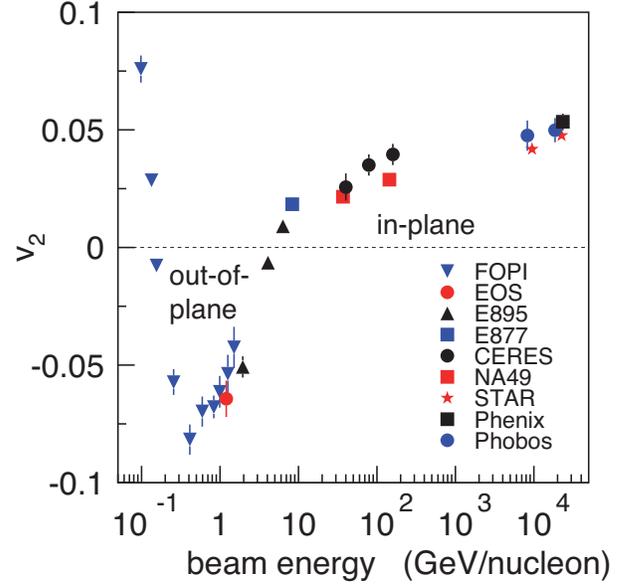}
\end{center}
\caption{(Color online) 
Elliptic flow $v_2$ of $Z=1$-particles at mid-rapidity as a function of
  incident beam energy in semi-central \AuAu collisions as measured by various
  experiments, indicated by the different symbols. Data are extracted from
  refs.~\cite{Andronic:2004cp,Andronic:2006ra,Danielewicz:2002pu,Pinkenburg:1999ya}
}
\label{FigAnton}
\end{figure}

The elliptic flow is described 
by the second  moment of the Fourier expansion $v_2$ of the azimuthal angle $\phi$
distribution of the emitted particles with respect to the reaction plane
$\Phi_{RP}$. All expansion coefficients $v_n$ are typically 
functions of rapidity $y=\frac{1}{2}\text{ln}\left(\frac{E+p_z}{E-p_z}\right)$ and 
of transverse momentum $p_t$ of the particle: 
\bea
\frac{d\sigma(y,p_t)}{d\phi}= C (1+ 2v_1(y,p_t) \cos{(\phi-\Phi_{RP})} \nonumber \\
+ 2v_2(y,p_t) \cos{2(\phi-\Phi_{RP})}+...)
\label{ellip}
\eea
The Fourier coefficients are then determined by:
\bea
\langle v_n(y,p_t) \rangle = \langle cos[n(\phi-\Phi_{RP})]\rangle \nonumber \\
\mbox{ with }  v_2  = \frac{p_x^2-p_y^2}{p_x^2+p_y^2}
\eea
where the angular brackets denote an averaging over all events and particles at
$y$ and $p_t$. A positive $v_2$-value characterizes a preferred emission in the reaction plane and a
negative value an emission out of the reaction plane. 
In Fig.\,\ref{FigAnton} experimental results of $v_2$ parameters for
$Z=1$-particles at mid-rapidity for  
semi-central Au+Au collisions at various energies are compiled.
%
%

At ultra-relativistic energies the measured elliptic flow and its
centrality dependence has been considered as an experimental proof that during
the expansion of the system the almond shaped initial spatial configuration of the overlap region
is transformed into an elliptic flow with a {\it positive} $v_2$ value as 
predicted by  hydrodynamics \cite{Luzum:2009sb}. At lower energies
various experimental groups \cite{Gutbrod:1989gh, Pinkenburg:1999ya} and later the FOPI
collaboration  \cite{FOPI:2011aa} have investigated the elliptic flow and found a
{\it negative} $v_2$ coefficient up to beam  energies of $\approx$ 6 AGeV with
a minimum at around 0.4-0.6 AGeV
\cite{Andronic:2004cp,Stoicea:2004kp,Andronic:2006ra}. 
Therefore, the elliptic flow has to be of different
origin at these energies. It has been suggested in \cite{Danielewicz:2002pu}
that the $v_2$ values are negative at low energies because the compressed
matter  expands while the spectator matter is still  present and blocks the
in-plane emission. At higher incident energies the expansion takes place after
the spectator matter has passed the compressed zone and therefore the elliptic
flow is determined by the shape of the overlap region only, which leads to a
positive $v_2$. The negative $v_2$ at
low incident energies is
due to shadowing overlaid by an expansion of the
compressed overlap zone \cite{Stoicea:2004kp}. The minimum  of the elliptic
flow $v_2$ coincides
with the maximum of nuclear stopping at these energies \cite{Reisdorf:2004wg} and high baryon
densities are reached during the collision. Contrary to findings at
higher beam energies where fluctuations contribute to the elliptic flow 
(see e.g. \cite{Ollitrault:2009ie,Pang:2012he})
there is no convincing experimental evidence at beam energies between 0.4
and 2A~GeV that event-by-event fluctuations play a significant role in the
elliptic flow pattern
\cite{Bastid:2005ct}. The interactions with the surrounding
spectator matter and the much longer collision times might be responsible
for this. 
At even lower incident energies $v_2$ becomes positive again,
because the attractive NN forces outweigh the repulsive NN collisions. This
phenomenon has been discussed in various publications,
e.g. \cite{Lukasik:2004df,  Zheng:1999gt,Sahu:1999mq}. 

Recently, the FOPI collaboration has compared its experimental findings on
elliptic flow $v_2$ of light charged particles measured in Au+Au collisions with
results obtained in the 
framework of Quantum Molecular Dynamics (QMD) calculations
\cite{Fevre:2015fza}. One conclusion was that the elliptic flow at energies
between 0.2 and 2.0 AGeV has 
the largest dependence on the stiffness of the nuclear EoS of all observables studied so far, 
an even larger dependence than found earlier in Kaon production
\cite{Hartnack:2005tr}. 
These findings created therefore a
renewed interest to study in detail the origin of the elliptic flow and its
dependence on the EoS. In this article we report on investigations using the
Isospin Quantum Molecular Dynamics model. In
section 2 we will shortly introduce the Quantum Molecular Dynamics (QMD)
approach which we use for the analysis. Section 3 is devoted to a survey of
the reaction, especially to a comparison of the reaction scenarios for
different EOS's. In section 4 we study in detail the elliptic flow created in
these reaction and analyze its origin and its EoS dependence. We summarize our
work in section 5.

\section{The Quantum Molecular Dynamics approach}

The details of the Quantum Molecular Dynamics (QMD) approach have been
published in \cite{Aichelin:1991xy,Hartnack:1997ez,Hartnack:2011cn}. 
Comparisons to experimental bench-mark data measured in the incident energy
region under consideration are published in \cite{FOPI:2011aa}.
Here, we quote only how this approach allows for an exploration of the
nuclear EoS:

Nucleons are represented as Gaussian wave functions. A
generalized Ritz variational principle allows to determine the time evolution
of the centroids of the Gaussians in coordinate $r_i$ and momentum space $p_i$. 

\begin{equation}
\dot{r_i}=\frac{\partial\langle H \rangle}{\partial p_i} \qquad
\dot{p_i}=-\frac{\partial \langle H \rangle}{\partial r_i} \quad ,
\end{equation}
where the expectation value of the total Hamiltonian $H$ is
\begin{eqnarray}
\langle H \rangle &=& \langle T \rangle + \langle V \rangle
\nonumber \\ 
&=& \sum_i \frac{p_i^2}{2m_i} +
\sum_{i} \sum_{j>i}
 \int f_i({\bf r, p},t) \,
V({\bf r, r\,', p, p\,'}) \nonumber \\
&\cdot &f_j({\bf r\,', p\,'},t)\, \rm d{\bf r}\, \rm d{\bf r\,'}
\rm d{\bf p}\, \rm d{\bf p\,'} \quad.
\end{eqnarray}
$f_i$ is the single-particle Wigner
density
\begin{equation} \label{fdefinition}
 f_i ({\bf r, p},t) = \frac{1}{\pi^3 \hbar^3 }
 {\rm e}^{-\frac{2}{L} ({\bf r} - {\bf r_i} (t) )^2   }
 {\rm e}^{-\frac{L}{2\hbar^2} ({\bf p - p_i} (t) )^2 } \quad .
\end{equation}
The potential consists of several terms:
\begin{eqnarray}
V({\bf r_i, r_j, p_i, p_j}) &=& G + V_{\rm Coul} \nonumber \\
       &=& V_{\rm Skyrme} + V_{\rm Yuk} + V_{\rm mdi} +
           + V_{sym} + V_{\rm Coul}  \nonumber \\
       &=& t_1 \delta ({\bf r_i} - {\bf r_j}) +
           t_2 \delta ({\bf r_i} - {\bf r_j}) \rho^{\gamma-1}({\bf r_i}) +
       \nonumber \\
       & &  t_3 \frac{\hbox{exp}\{-|{\bf r_i}-{\bf r_j}|/\mu\}}
               {|{\bf r_i}-{\bf r_j}|/\mu} +  \\
       & & t_4\hbox{ln}^2 (1+t_5(\bf{p}_i-\bf{p}_j)^2)
               \delta ({\bf r_i} -{\bf r_j}) +\nonumber \\
       & & t_6 \frac{1}{\varrho_0}
                T_{3}^i T_{3}^j \delta({\bf r_i} - {\bf r_j}) + 
        \frac{Z_i Z_j e^2}{|{\bf r_i}-{\bf r_j}|} \nonumber .
\label{Vij}
\end{eqnarray}

The total one-body Wigner density is the sum of the Wigner densities of
all nucleons.
The nuclear EoS, on the other hand, describes
the properties of infinite nuclear matter (without Coulomb
interactions) and is therefore given by the volume energy only.
The EoS describes the variation of the energy
$E(T=0,\rho/\rho_0)$
when changing the nuclear density to values different from the saturation
density $\rho_0$ for zero temperature.

The single-particle potential resulting from the convolution of the distribution
functions $f_i$ and $f_j$ with the interactions
$V_{\rm Skyrme}+ V_{\rm mdi}$
(local interactions including their momentum dependence) is for symmetric nuclear matter
\begin{equation} \label{eosinf}
U_i({\bf r_i},t) \,=\, \alpha
\left(\frac{\rho_{int}}{\rho_0}\right) +
        \beta \left(\frac{\rho_{int}}{\rho_0}\right)^{\gamma} +
        \delta \mbox{ln}^2 \left( \varepsilon
                \left( \Delta {\bf p} \right)^2 +1 \right)
                        \left(\frac{\rho_{int}}{\rho_0}\right) \quad ,
\label{Upot}
\end{equation}
where $\rho_{int}$ is the interaction density obtained by
convoluting the distribution function of a particle with the
distribution functions of all other particles of the surrounding
medium. $\Delta {\bf{p}}$ is the relative momentum of a particle
with respect to the surrounding medium.

In nuclear matter the parameters $t_1, t_2, t_4, t_5$ in Eq.~\ref{Vij} are uniquely
related to the coefficients $\alpha, \beta, \delta$, and
$\epsilon$ in Eq.~\ref{Upot}. Values of these parameters for the
different model choices can be found in Tab.~\ref{eostab}.

\begin{table}[hbt]
\begin{tabular}{lcccccc}
 &$\alpha$ (MeV)  &$\beta$ (MeV) & $\gamma$ & $\delta$ (MeV) &$\varepsilon \,
 \left(\frac{c^2}{\mbox{GeV}^2}\right) $ & $K$ (MeV) \\
\hline
 SM & -390 & 320 & 1.14 & 1.57 & 500  & 200 \\
 HM & -130 & 59  & 2.09 & 1.57 & 500  & 376 \\
\end{tabular}
\caption{Parameter sets for the nuclear equation of state used in
the IQMD model} \label{eostab}
\end{table}

The parameters $\epsilon$ and $\delta$ are given by fits to the
optical potential extracted from elastic scattering data in pA
collisions\cite{Hama:1990vr,Aichelin:1987ti}.  
Two of the 3 remaining  parameters of the ansatz are fixed by the condition that the
volume energy has a minimum of $E/A(\rho_0)=-16$ MeV at $\rho_0$.

The third parameter is historically expressed as the compression modulus $K$
of nuclear matter, which corresponds to the curvature of the 
volume energy at $\rho=\rho_0$ (for $T=0$) and is also given in
Tab. I
\begin{equation}
K  = -V\frac{{\rm \partial}p}{{\rm \partial}V}= 9 \rho^2
\frac{{\rm \partial}^2E/A(\rho)}{({\rm \partial}\rho)^2} |_{\rho=\rho_0} \qquad
.
\end{equation}

An equation of state with a rather low value
of the compression modulus $K$ yields a weak repulsion of
compressed nuclear matter and thus describes "soft" matter
(denoted by "SM"). A high value of $K$ causes a strong repulsion of
nuclear matter under compression (called a "hard EoS", HM). 

Generally, there is a good agreement between the model results and
experimental data. Critical input parameters, like cross sections, are confined by
experimental observations. Shortcomings of the model are pion production
and the formation of heavy clusters at mid-rapidity, which are discussed in
detail in reference \cite{Fevre:2015fza}.

\section{Survey of the reaction}

Motivated by the good agreement between experimental data and the results of
the IQMD model in most of the relevant
flow observables \cite{FOPI:2011aa,Fevre:2015fza}, we use this
model in order to understand the reaction in its full complexity.
\AuAu collisions at 0.6 and 1.5~AGeV and an impact parameter of 6 fm 
are used as model cases, because at around 0.6
AGeV the elliptic flow excitation function reaches its minimum and 1.5~AGeV
is the highest energy measured by the FOPI collaboration.  For the
following discussion, only protons were taken into
consideration. We verified that neither the formation of clusters nor the behavior of
neutrons alter our findings.

\begin{figure}[t]
\centering
\includegraphics[width=\columnwidth]{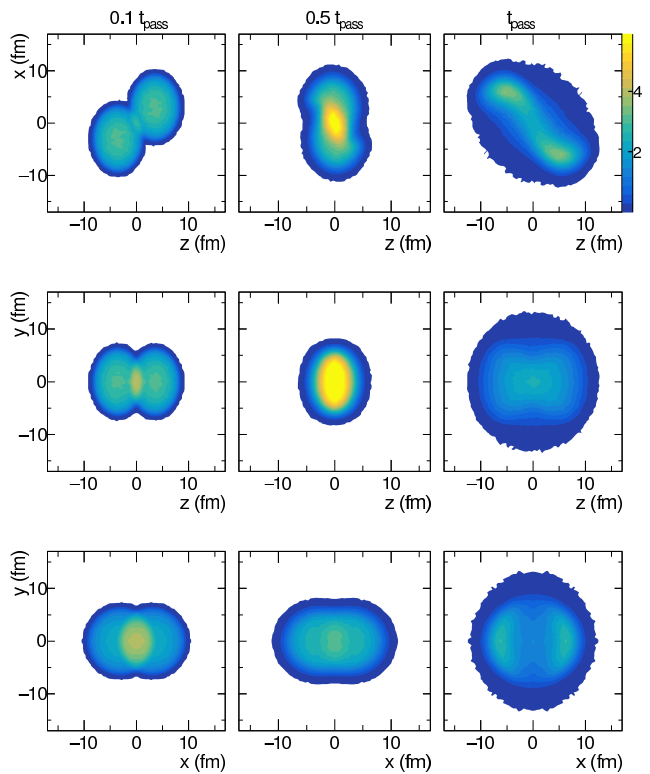}
  \caption{(Color online) Time evolution of the proton density profile
    ($\rho_{ij}$ in $event^{-1}fm^{-2}$) in the reaction \AuAu at 1.5 AGeV
    incident energy with an impact parameter b = 6 fm. A soft (SM) EoS is
    applied in the calculations. Different projections are shown: projections
    onto the zx-plane in the top row, onto the zy-plane in the middle, and
    onto the xy-plane in the bottom. 
The density profile are shown at different times: 2, 8 and 16 fm/c (0.1, 0.5, 1.0
$t_{pass}$) in the left, central and right columns, respectively.} 
\label{FigNxyz1500}
\end{figure}

The time evolution of the heavy ion reaction
\AuAu at $E_{kin}= 1.5$~AGeV  
and an impact parameter b=6~fm is shown in Fig.~\ref{FigNxyz1500}. The density profile of protons,
($\rho_{ij}=\frac{1}{N_{event}}\int\frac{dN_p}{didjdk}dk$, where i,j,k
represent the three space coordinates x, y, z)   
at different times expressed in units of the passing time  $t_{pass}$ is presented. The passing
time,  $t_{pass}$, is the time
the nuclei need to pass each other completely assuming that they do not
experience deceleration and therefore continue moving with their initial velocity. For
\AuAu collisions at $E_{kin}= 0.6$~AGeV the passing time is $t_{pass} = 22.9$~fm/c
and 16.9~fm/c for $E_{kin}= 1.5$~AGeV.
After $t_{pass}$ the spectator matter (those nucleons of projectile and target which are
outside of the overlap of projectile and target) cannot absorb nucleons
from the participant region (the nucleons of the overlap region of projectile
and target) anymore. Projections of proton densities $\rho(x,y,z)$ onto the zx plane (where
x is the direction of the impact parameter and z the direction of the beam),
onto the zy plane and onto the xy plane are shown from top to bottom and for
three different times $t=0.1, 0.5, 1.0 t_{pass}$ from right to left. 
As can be seen in the top figures, the central (participant) matter
is highly compressed when the overlap of the colliding system is largest 
at $t=0.5 t_{pass}$. Projectile and target remnants separate but
they are connected for longer than $t_{pass}$ 
by a ridge with a quite high particle density. This ridge will
disintegrate when projectile and target remnants separate further. The
importance of this ridge can be seen in the second row which shows 
the density profile in the zy plane. In this projection, at half $t_{pass}$, we observe the highest
density at z=0 and therefore in the ridge. 


\begin{figure}[ht]
\centering\includegraphics[width=\columnwidth]{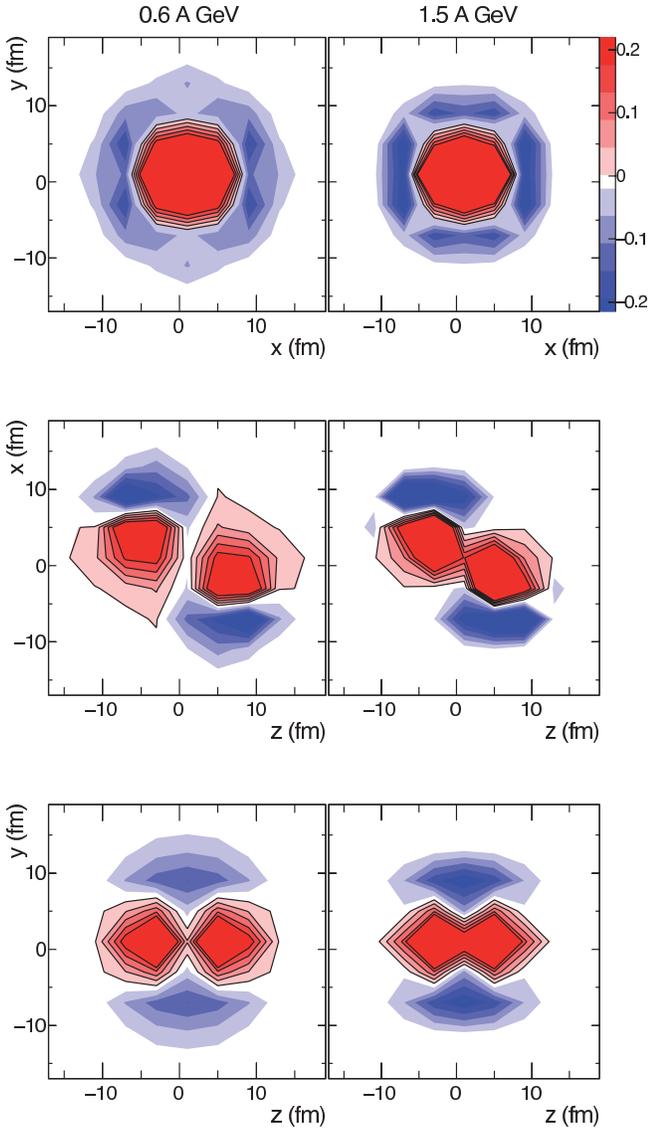}
\caption{(Color online) Difference of the proton density profiles,
$\delta\rho_{x,y} = \rho_{xy}{}^{SM}-\rho_{xy}{}^{HM}$ in $event^{-1}fm^{-2}$) between a soft
  (SM) and a hard (HM) EoS in the (from top to bottom) xy, zx and zy
  planes. Predictions for \AuAu at 0.6 AGeV  and 1.5 AGeV incident energies
  are shown on the left and columns respectively. The impact parameter is
  6\ fm and the model results are for $t=t_{pass}$. The red color stands for
  positive values, the blue color for negative ones. Positive values are
  emphasized with black contour lines in addition.
} 
\label{FigdNxyzSMHM}
\end{figure}

\begin{figure}
\centering
\includegraphics[width=\columnwidth]{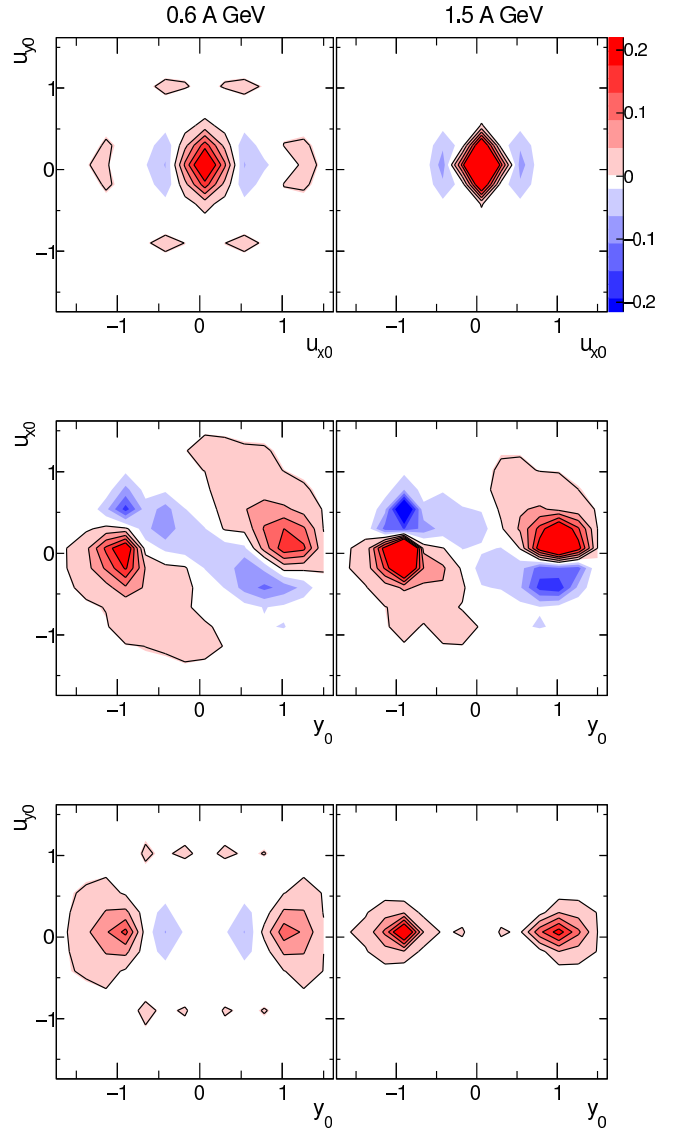}
\caption{(Color online) Same as Fig. \ref{FigdNxyzSMHM} but expressed in the
  planes of $(u_{x0},u_{y0})$, $(u_{x0},y_{0})$,
  $(u_{y0},y_{0})$ scaled velocities (see text).}  
\label{FigdNpxyzSMHM}
\end{figure}

 \begin{figure} 
  \centering
  \includegraphics[width=0.9\columnwidth]{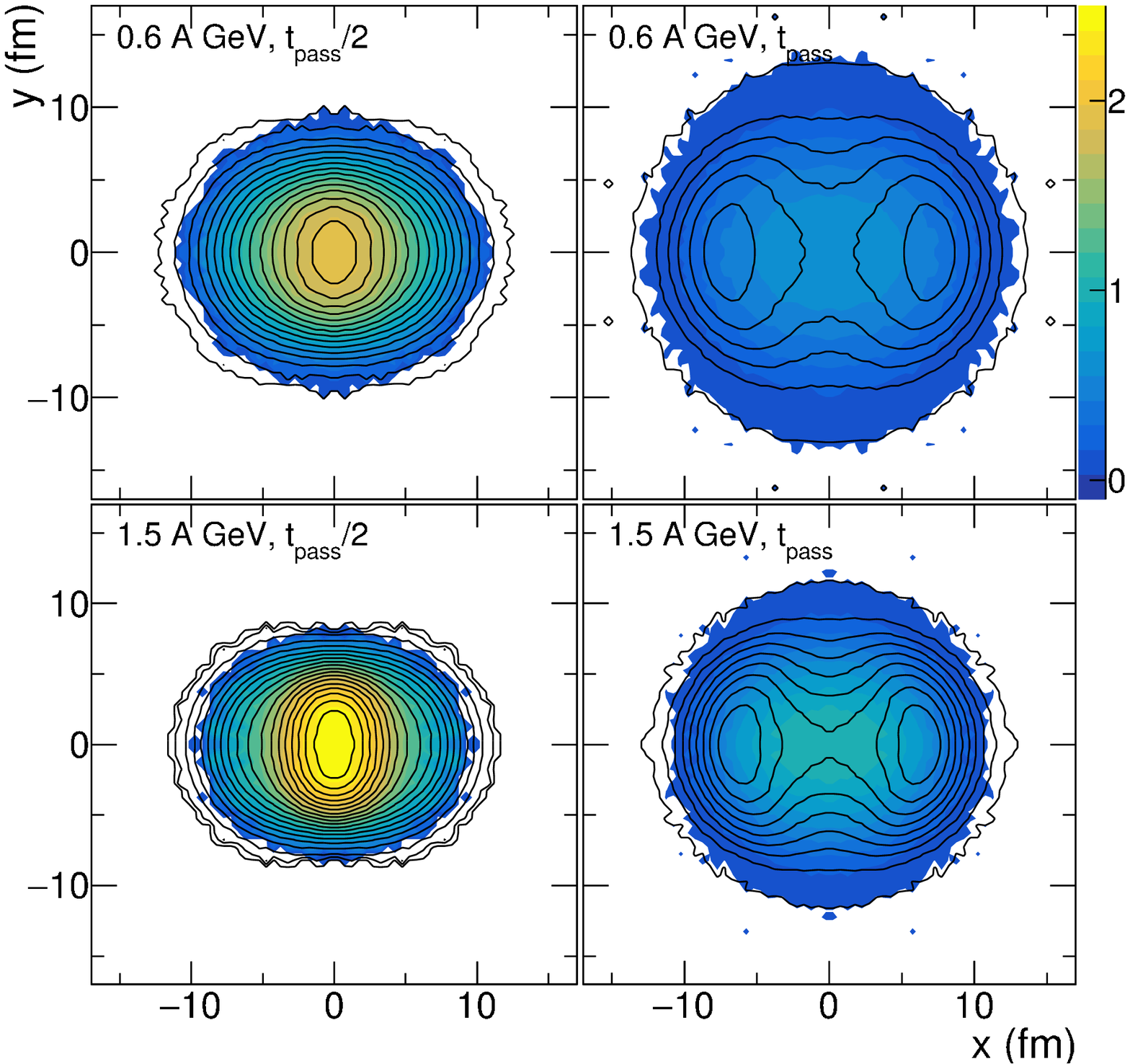}
  \includegraphics[width=0.9\columnwidth]{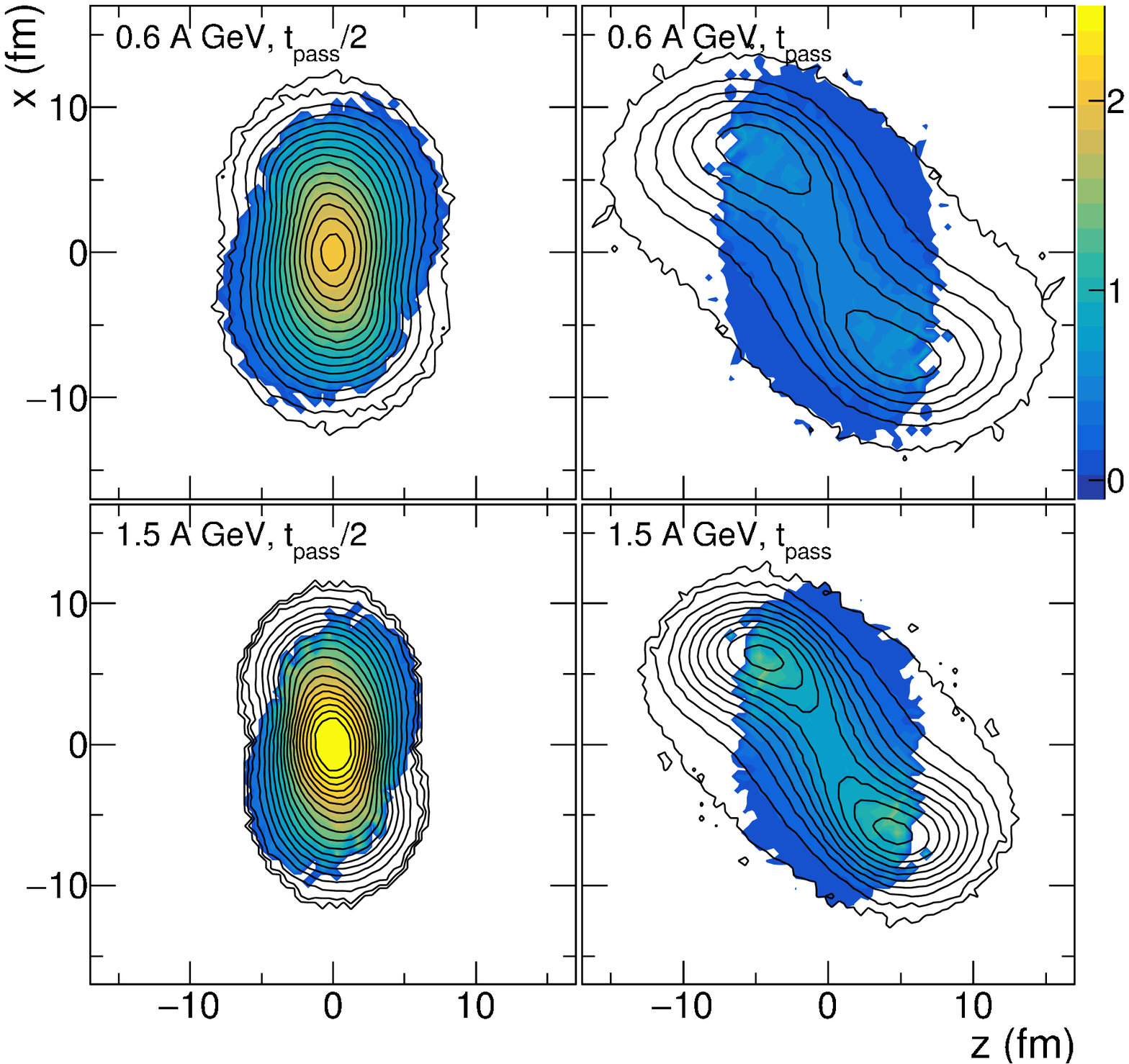}
 \caption{(Color online) Mean reduced
    baryonic density ($\rho/\rho_0$) in coordinate space as perceived by
    protons in \AuAu collisions with a
    soft (SM) EoS,  b=6 fm,  at 0.6 (top) and 1.5 (bottom) AGeV 
    incident energies, at two different times: at full overlap of the
    system $0.5 t_{pass}$ (left) and at the passing time $t_{pass}$ (right). 
    Black lines and colored contours correspond to all protons and to those 
    finally emitted at mid-rapidity ($|y_0|<0.2$)  with a high
    transverse velocity ($u_{t0}>0.4$), respectively. 
The top four-panel groups show projections on xy plane, and the lower ones
projctions on the xz planes.
} 
\label{fig:fig14}
\end{figure}

The choice of the EoS influences the reaction scenario predicted by the
model. This can be studied in detail by evaluating the
difference  (SM-HM) of the proton densities projected onto the xy plane, 
$\Delta
\rho_{xy}=\rho_{xy}^{SM}-\rho_{xy}^{HM}$,   
and correspondingly onto the zx and zy  planes. The results
are shown in Figs.~\ref{FigdNxyzSMHM} and \ref{FigdNpxyzSMHM} in coordinate
and velocity space, respectively. The red color
signals regions in which a soft EoS yields a higher density, whereas  the blue
color marks the regions in which the density is higher for a hard EoS. In
addition, the positive values are emphasized with black contour lines.   
Fig. \ref{FigdNxyzSMHM}
(\ref{FigdNpxyzSMHM}) displays this density difference at $ t_{pass}$ 
in coordinate (velocity) space for the reaction \AuAu at 0.6~AGeV (left) and
at 1.5~AGeV (right).
The density of protons in the
geometrical overlap region of projectile and target is substantially higher 
for a soft EoS, 
as can be seen in the uppermost panels of Fig. \ref{FigdNxyzSMHM}, 
whereas at larger distances
from the reaction center we observe a higher density for a hard EoS. 
At $0.6$ AGeV this surplus in the density for a hard EoS in the xy plane
 is larger in x-direction, but it becomes rather isotropic  at 1.5 AGeV.  
The origin of this surplus in x-direction is rather different
from that in y-direction: in the middle panel of Fig.~\ref{FigdNxyzSMHM}  is shown
that the excess in x-direction has its origin in the in-plane flow of the
spectator matter expressed by a finite $v_1$ coefficient in Eq. \ref{ellip}. This
in-plane flow is considerably stronger for a hard EoS as compared to a soft
one \cite{Aichelin:1987ti,Peilert:1989kr,Hartnack:1999gn}. In y-direction the
surplus in density of the hard EoS is concentrated at 
around z=0, being less extended but stronger at higher energies 
(Fig. \ref{FigdNxyzSMHM} lower panels). 
The emission of these particles is caused by a stronger density gradient
(and hence a stronger force) in y-direction for a hard (HM) EoS 
as compared to a soft (SM) one.

In order to analyze the model results in momentum space 
we introduce the transverse vector $ \frac{\vec{p_t}}{m}= \vec{u_t} = \vec{\beta_t}
\gamma_t$ with $\vec{\beta_t} = (\beta_x,\beta_y)$. The 3-vector $\vec {\beta} $  is the velocity in units of the light
velocity and $\gamma = 1/\sqrt{1-  \beta^2}$. Throughout, we use  
scaled units for the rapidity $y_0 = y/y_p$ and the transverse velocity $\vec{u_{t0}} = (u_{x0},u_{y0}) = \vec{u_t}/u_p$, with $u_p =
\beta_p\gamma_p$, the index p referring to the 
incident projectile in the center of the colliding system. In these units the initial
target-projectile rapidity gap always extends from $y_0 = -1$ to $y_0 = 1$. 

In velocity space (Fig. \ref{FigdNpxyzSMHM}) we observe a complementary
distribution. In the xy plane (upper panels) the shift
of protons in x direction is smaller for a soft (SM) than for a hard (HM) EoS 
due to a smaller acceleration yielding a weaker in-plane flow and hence a smaller
velocity in x-direction (see middle panels).
The soft EoS leads also to less stopping, as can be seen in the lower panels.

Fast moving particles in the transverse direction at 
mid-rapidity are selected by applying the following cuts: $|y_{0}|<0.2$, $u_{t0}>0.4$. Identical
cuts were used by the FOPI collaboration for the investigation of elliptic
flow.
Fig. \ref{fig:fig14} shows the averaged normalized nuclear density
($\rho/\rho_0$) obtained for a soft (SM) EoS for this selection of participant protons
in the xy plane (upper four panels) and in
the zx plane (lower four panels) for 0.6 and 1.5 AGeV incident energies at
$t=0.5 t_{pass}$ (left column) and $t=t_{pass}$ (right column). 
The density profiles are integrated over the third dimension. We confront
this average density (color
scaled) of protons finally observed with high velocities at mid-rapidity with
the density of all protons (contours). We observe that at full
overlap, $t = 0.5 t_{pass}$, 
the innermost participants form a dense almond shaped core which is
out-of-plane elongated. This is the target-projectile overlap region, 
where the compression is highest.
On the contrary, the outermost participants, which form a more dilute medium, are extending in-plane, 
aligned with the spectator distribution, though slightly tilted as a consequence of stopping. 
Later, at passing time (right panel), the innermost (compressed) participants
expand in-plane, but not with enough pressure to produce a positive elliptic flow $v_2$, as we
will see later. This is in contrast to the situation at higher bombarding energies
where the strength of the compression is high enough to make the in-plane expansion dominant. 
The outermost participants undergo a twofold evolution: First by expanding out-of-plane
(seen on the xy plane) which will produce a negative $v_2$ as will be shown later. 
Second by forming an in-plane ridge between the bulk of the spectators (seen on the xz plane).
The higher the incident energy the higher is the density of this ridge and of the initial almond shaped core.

\section{The elliptic flow}

\begin{figure}
  \centering
  \includegraphics[width=0.9\columnwidth]{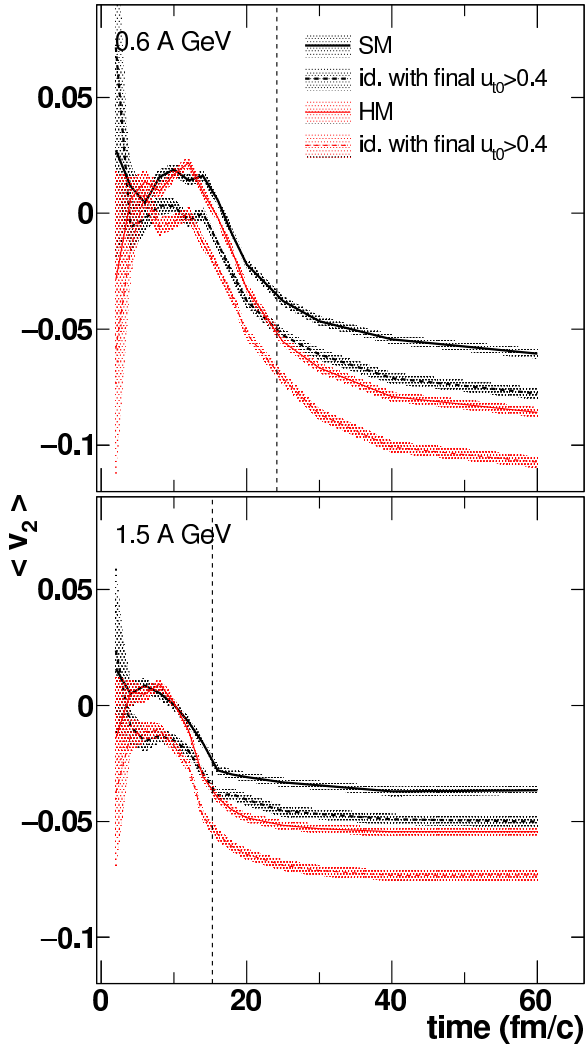}
  \caption{(Color online) Time evolution of the average elliptic flow $v_2(t)$
    of protons at midrapidity in the \AuAu collisions at 0.6 (top) and 1.5 (bottom) AGeV 
    incident energies, $b=6 fm$. We show results obtained with a hard (HM, red lines) and
    a soft (SM, black lines) EoS, and with (dashed lines) or without (full
    lines) excluding the protons having finally a low 
    transverse velocity $u_{t0} \leq 0.4$.
    The dashed vertical lines indicate the passing time, and the grayed areas
    the statistical uncertainties.} 
\label{FigV2time}
\end{figure}

\begin{figure}
\centering
\includegraphics[width=\columnwidth]{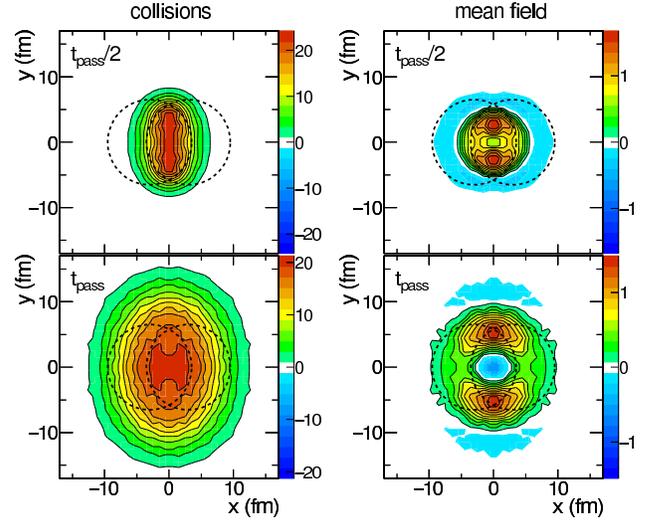}
  \caption{(Color online) IQMD (with SM EoS) predictions for mid-central (b=6 fm) collisions of
      \AuAu at $0.6$ AGeV incident energy, at two times: $t_{pass}/2$ (maximal
      overlap) and $t_{pass}$, top and bottom panels respectively. 
      The panels display $\Delta P_t^o(t)$ in
      $MeV/fm^2/event$ defined in the text as a function of the (x,y) positions of protons
      at the respective times. 
Only protons finally at mid-rapidity ($|y_0|<0.2$) are selected. 
The left and right panels show the momentum transfer due to collisions 
and to the mean field, respectively. As a reference, the superimposed circles show the  spatial
      extension of the incoming projectile and target in this plane. Positive
      values are marked by black contour lines. }
\label{fig:fig7}
\end{figure}

\begin{figure}
\centering
\includegraphics[width=\columnwidth]{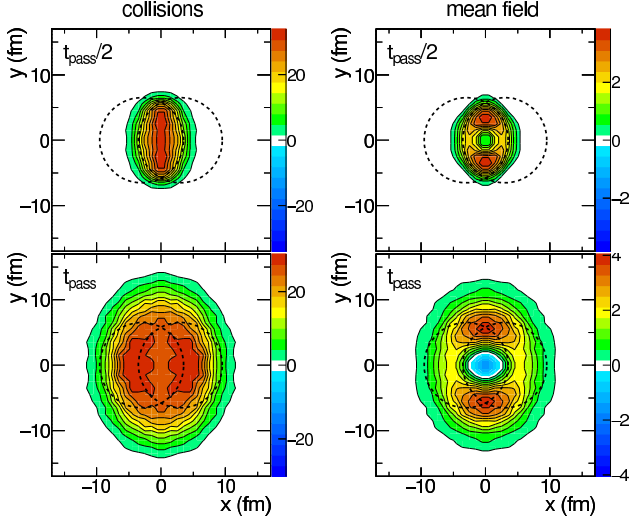}
  \caption{(Color online) Same as Fig. \ref{fig:fig7} for 1.5 AGeV incident energy.}
\label{fig:fig7bis}
\end{figure}

Fig.~\ref{FigV2time} shows the time evolution of the elliptic flow
$v_2(t)=\frac{p_x^2(t)-p_y^2(t)}{p_x^2(t)+p_y^2(t)}$  of mid-rapidity protons in the
\AuAu collisions at 0.6 (top) and 1.5 (bottom) AGeV for SM (black) and HM
(red) nuclear equation of state. 
The elliptic flow $v_2$  starts to develop after approximately half the
passing time $t_{pass}$ and evolves rapidly. After twice the passing time,
$v_2$ reaches its final value. It is negative for most of the collision times
and for both energies. But there is a tendency to be positive in the early stage of
the collision. If one selects  protons emerging with a high transverse velocity
$u_{t0}>0.4$ (dashed lines) the amplitude of the elliptic flow signal is
enhanced and it is mostly negative throughout the whole collision process. Comparing
the predictions for a soft (SM) and a hard (HM) equation of state one notes
that the value of $v_2$  at
mid-rapidity depends strongly on the EoS; this effect is enhanced if
 protons with a high transverse velocity are selected.

Scattering of nucleons and the mean field (potential) interactions are contributing to the
elliptic flow signal. In the simulations, it is possible to distinguish both
contributions and investigate how they develop as a function of time. 
This is achieved by recording the momenta of protons before and after each
collision and before and after each time step during which the proton
propagates in the potential created by all other nucleons. 

Hence, the momentum change due to collisions can be written as:

\begin{eqnarray}
\label{equ-coll}
\bm{{\Delta  P}{}^{coll}}(t) & = & \bm{p^{coll}}(t) -
\bm{p}(0) \nonumber \\
& = &\sum_{i=1}^{N_c(t)} \bm{\Delta p_i^{coll}} 
\end{eqnarray}
where $N_c(t)$ is the number of collisions a particle experiences until time $t$,
$\bm{\Delta p_i^{coll}}$ the momentum transfer in the $ i^{th}$
collision, and $\bm{p}(0)$ the initial momentum of the particle.

For potential interactions the time integrated momentum change is: 
\begin{eqnarray}
\label{equ-pot}
\bm{\Delta P^{m.f.}}(t) & = & \bm{p^{m.f.}}(t) -
\bm{p}(0) \nonumber \\
 & = & \sum_{i=0}^{i(t)} \int\limits^{t_{i+1}}_{t_i} \dot{\bm{p}}^{m.f.} dt 
\end{eqnarray}

With these prescriptions we define the momentum change into the transverse
direction as follows:
\begin{equation}
\bm{\Delta P_t{}^{coll,m.f.}}(t) = ( \Delta P_x{}^{coll,m.f.}(t), \Delta
P_y{}^{coll,m.f.}(t) ) 
\end{equation}
In order to visualize the effect of the momentum transfers on the elliptic
flow phenomena more strongly, we project the transverse momentum transfer
vector onto the final momentum vector of the
particle $\bm{p}_{final}$   

\begin{equation}
 \langle \Delta P_t^{o}(t) \rangle = \langle \bm{\Delta P_t}(t) 
\cdot \frac{{\bm p_{final}}}{|{\bm p_{final}}|} \rangle
\end{equation}

The angular brackets denote an averaging over events and particles.

Figs. \ref{fig:fig7} and \ref {fig:fig7bis} show this {\it oriented} transverse momentum
change $ \langle \Delta P_t^o(t) \rangle$ for beam energies of $0.6$\,AGeV and
$1.5$\,AGeV, respectively, 
separately for transverse momentum changes due to collisions (left panel)
and due to potential interactions (right panel) at different collision
times. The positive values are highlighted by black contour lines.
$\langle \Delta P_t^o(t) \rangle$ of protons
is shown as a function of their ($x(t)$, $y(t)$) position at half passing time $t=t_{pass}/2$ (top) and
at passing time $t=t_{pass}$ (bottom). Protons are selected which are
finally emitted at mid-rapidity ($|y_0|<0.2$).  
Comparing the scales of the left and right panels, one first observes that the
transverse momentum transfer due to collisions is about an order of magnitude  
larger than that due to potentials.  

In the overlap zone of projectile and target, where the number of collisions
is highest, the collisions create quite early (at half passing time) a large
value of $ \langle \Delta P_t^o(t) \rangle$. This means that 
the momentum transfer is large in the initial violent collisions
and the direction of  the particle momentum is - on the average - already
close to the final one. Because the nucleons gained a
considerable transverse momentum, this zone of violent collisions expands
rapidly keeping its almond shape.   

$ \langle \Delta P_t^o(t) \rangle $ due to potential interactions shows a quite
different structure.  The out-of plane 
momentum transfer is large in the vicinity of  the tips of the almond shape
overlap zone because these nucleons are directly situated between vacuum and
the central densest zone. Therefore they feel the highest density gradient and
hence the largest force. The comparison of the top (half passing time) and
bottom rows (passing time) shows how these accelerated particles move in
y-direction out of the overlap zone. Qualitatively there is little difference
between the reaction at 0.6 AGeV and at 1.5 AGeV. Particles distant from the
center of the reaction show a negative $ \langle \Delta P_t^o(t)
\rangle$. They are feeling 
the attractive potential of the remnant and are getting decelerated.  There is also a
zone around the origin where $ \langle \Delta P^o_t(t) \rangle $ is negative.  As Fig.~\ref{FigNxyz1500}
top right shows, these nucleons form the ridge between projectile and target
remnant. The density of the ridge around $z=x=y=0$ decreases between
$t_{pass}/2$ and $t_{pass}$. But the nuclear matter is attracted by the moving
spectators in the xz-plane and its velocity in transverse direction is
reduced.

\begin{figure*}
\centering
\includegraphics[width=0.65\textwidth]{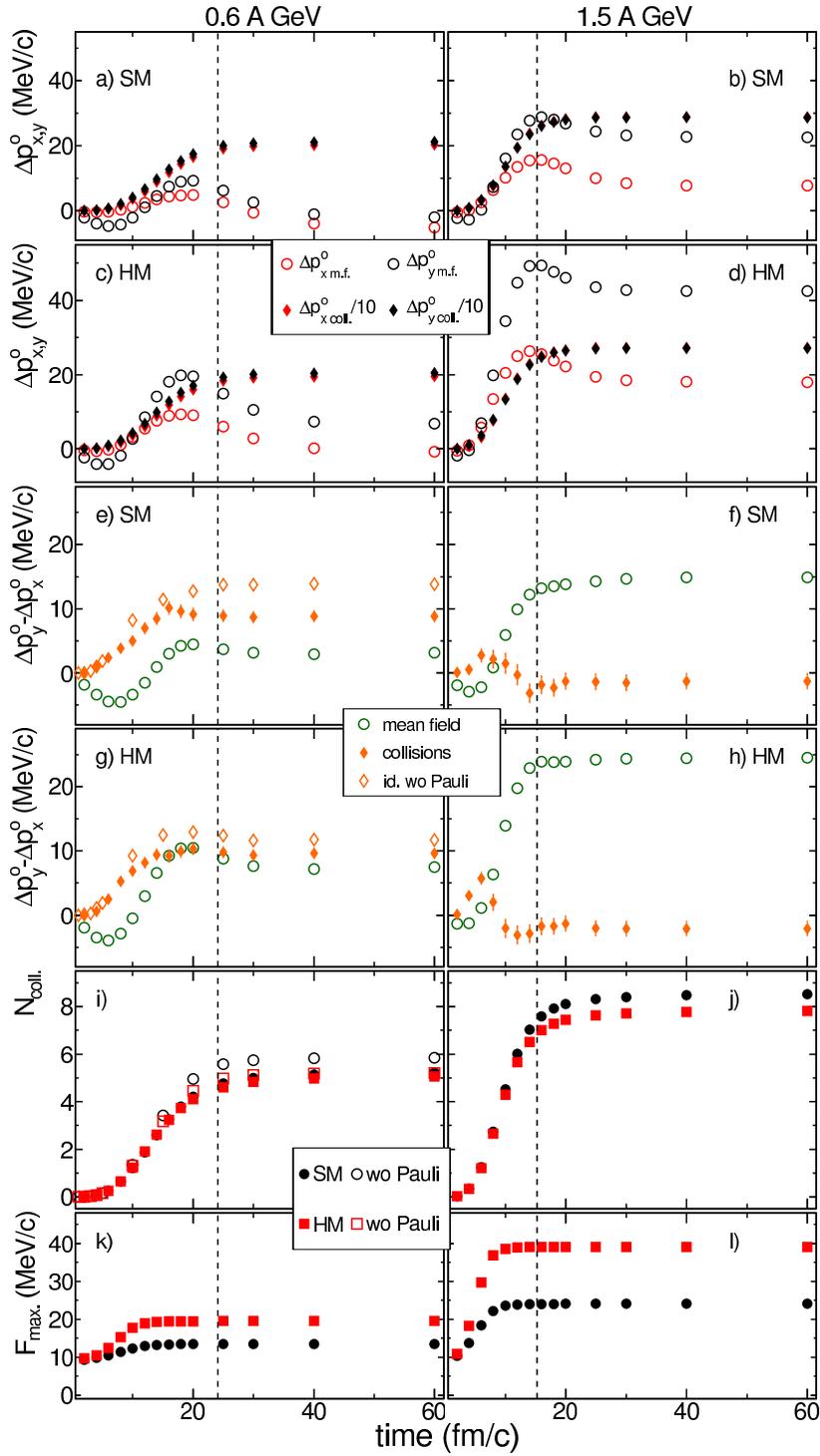}
  \caption{IQMD predictions of the time evolution of various
    observables. Protons are selected which are finally  observed at
    mid-rapidity ($|y_0|<0.2$) in the mid-peripheral (b=6 fm) collisions of
    \AuAu.
    Left and right panels show results at 0.6 and 1.5 AGeV incident energies, respectively.  
    Panels a, b: show for the SM EoS the 
    integrated momentum transfer in x (red
    symbols) and y (black symbols) directions, caused by 
    the mean field (m.f., open circles) and by collisions (coll.,
    diamonds). The momentum transfer due to collisions is divided by a factor
    of 10.
    Panels c and d: \textit{idem} for the HM EoS.
    Panels e and f: show for the  SM EoS the integrated difference
    between the out-of-plane (y) and in-plane (x) contribution of the
    momentum transfer from the mean field (green open circles) and
    from the collisions with (orange full diamonds) or without (orange
    empty diamonds) Pauli blocking. 
    Panels g and h: \textit{idem} for the HM EoS.
    Panels i and j: number of
    collisions suffered by the selected protons comparing the SM (black circles) and HM (red squares)
    EOS's, with (full symbols) or without (open symbols) Pauli blocking.
    Panels k and l: \textit{idem} with the maximal force due to the
    mean field.
    The vertical dashed lines indicate the passing time. 
}
  \label{fig:fig3}
\end{figure*}

The elliptic flow $v_2$ is not related to the magnitude of the transverse
momentum change $ \langle \Delta P^o(t) \rangle$ but to its anisotropy in x and y. 
To access this situation the quantity  $\Delta P_{y-x}^o(t)  
=\Delta P_y^o(t) - \Delta P_x^o(t)$ is introduced. For a single proton the
directed momentum 
change ${\Delta P_i^o(t)}$ is defined by the momentum change
in x or y-direction ${\Delta P_{i}(t)}$ projected onto the direction of the
respective component of the final momentum vector, 
\begin{equation}
\langle \Delta P_i^o(t) \rangle  = \langle \Delta P_i(t) \cdot
\frac{p_{i,final}}{|p_{i,final}|} \rangle .
\end{equation} 
$\langle \Delta P_i^o \rangle$ is calculated for momentum changes due to potentials  and due to
collisions defined in equations \ref{equ-pot} and \ref{equ-coll}, respectively.

The resulting quantities are presented in Fig. \ref{fig:fig3}, where results of model calculations
are shown for protons emitted at mid-rapidity, $|y_0|<0.2$, in
\AuAu collisions of 0.6 (right panels) and 1.5~AGeV (left panels) at impact
parameter $b = 6$~fm.
Panels a), b), c), d) show the time dependence of the momentum change $\Delta
P_x^o(t)$ and $\Delta P_y^o(t)$
integrated up to time $t$ due to mean field interactions (black symbols) and
due to collisions (red symbols) for different nuclear equations
of state. The integrated momentum change due to collisions is
always much larger than the one generated by the mean field. Note, that
in this figure the data for momentum changes due to collisions are divided by a factor
of 10. For this type of momentum changes 
one observes only a rather small excess in the y-direction
(out-of-plane) at the low energy and essentially none 
at the high energy. But an excess in the y-direction is always visible
for the momentum changes due to potential interactions. 
This is quantified in panels e), f), g), and h) where the difference 
$\Delta P_y^o(t)-\Delta P_x^o(t)$ is
presented as a function of time. 
The excess in the y-direction is clearly visible for the potential interaction,
but also the collisions produce such an effect
with an amplitude which  becomes smaller with higher projectile velocity until it vanishes
at  $1.5$ AGeV incident energy. 

The stiffness of the equation of state has no visible influence on the amplitude of the collisional
out-of-plane  momentum excess. 
This is related to the fact that the number of collisions, as displayed in panels i)
and j), is almost unchanged by the choice of the equation of state. The correlation of 
the time evolution of the collisional $\Delta P_y^o-\Delta P_x^o$ with the number of collisions 
is particularly marked at the lower incident energy (panel i).

An additional reason for the EoS independence of the collisional out-of-plane
flow is the Pauli blocking. Its influence is only studied for the lower energy
because it is negligible at the higher one. 
The open red diamonds in panels e) and g) show the effect on collisional $\Delta
P_y^o-\Delta P_x^o$ when switching-off Pauli blocking. Without Pauli blocking there is a
visible sensitivity to the nuclear EoS for this observables when only 
collisional contributions are considered.
However, Pauli blocking quenches the out-of-plane flow due to collisions, starting from the densest
phase of the collisions, at half $t_{pass}$ (maximal overlap). 
The quenching is stronger for the softer (SM) EoS because the
central hadron densities reached during the collision process are larger.  
Thus, the model predicts that without Pauli blocking there would be a collisional contribution to the
EoS dependence of  $v_2$, but with Pauli blocking this
sensitivity is vanishing, which finally leads to the observation
that there is no collisional contribution to the EOS dependence of the $v_2$ signal.
 
The mean field contribution to the out-of-plane momentum flow is enhanced by
both, the incident energy and the  
stiffness of the equation of state: moderate at $0.6$ AGeV with the soft (SM)
EoS, contributing to only $30\%$ of the  
total $\Delta P_y^o-\Delta P_x^o$, very strong and dominating at $1.5$ AGeV with the stiffer (HM) EoS. 
This is directly correlated with the strength of the mean field, displayed in panels k and l,  
nearly doubled for the harder (HM) EoS. In conclusion, we observe that the only essential dependence 
of the out-of-plane flow on the EOS  comes from the mean field. 

\begin{figure} 
  \centering
  \includegraphics[width=\columnwidth]{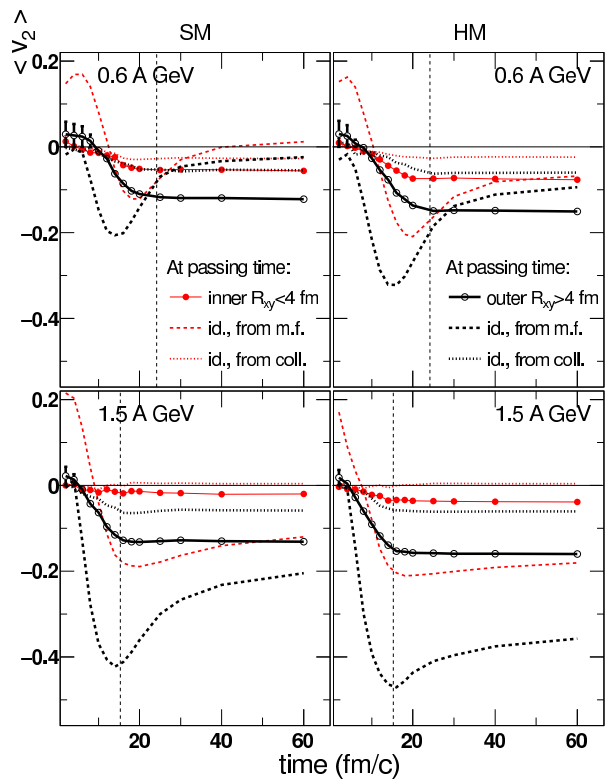}
\caption{(Color online)
Time evolution of the average elliptic flow, $v_2$, of protons finally emitted at mid-rapidity ($|y_0|<0.2$) 
with a large transverse velocity $u_{t0}>0.4$ 
in \AuAu collisions at b=6 fm and  at 0.6 (top) and 1.5 (bottom) AGeV
incident energy, with a soft (SM, right) and a hard (HM, left)
EoS. The protons situated, at the passing time, transversally close
(radial distance to the center
of the collision on the transversal plane $R_{xy}<4 fm$) or far
($R_{xy}>4 fm$) to/from the main axis of the collision 
 are distinguished, respectively, by red and
black lines. The overall $v_2$ (symbols) is detailed into its two
contributions: the $v_2$ developed by the momentum transfer due to the 
mean field and due to the collisions are depicted by dashed and
dotted lines, respectively. Black vertical dashed lines indicate the passing time.
}
\label{fig:fig16}
\end{figure}

The origin of $v_2$ is further investigated by
analyzing the elliptic flow in the xy plane as a function of the transversal
distance of the protons from the center of the reaction. The positions of the
protons are evaluated at $t_{pass}$. The results of such an analysis 
are presented in Fig. \ref{fig:fig16}.
As before, protons were selected which are finally emerging in the
mid-rapidity region $|y_{0}|<0.2$ with high transverse velocity $u_{t0}$ which only enhances the amplitude
of the observed phenomena, as discussed in Fig.~\ref{FigV2time}. 
First we observe that the collisional contribution to $v_2$ reaches its asymptotic value early,
before or close to the passing time $t_{pass}$, when collisions cease, 
as seen in Fig. \ref{fig:fig3} panels i and j. The collisional contribution to
$v_2$ is a fast process because it needs the presence of the spectators 
to induce an in-plane quenching effect. 
The mean field contribution 
stabilizes at a slightly later time at 0.6 AGeV and even later at 1.5 AGeV, long
after the strength of the force, shown in Fig. \ref{fig:fig3} panels k and l,
reaches its maximal value.  
 
Another feature is that the outermost nucleons ($R_{xy}>4 fm$) are the
main source of the overall negative $v_2$, they 
develop a much stronger out-of-plane flow.  This is observed for the
collisional contribution because the early in-plane screening by the
spectators affects only the outermost nucleons, 
whereas the collisions of the inner nucleons create a nearly azimuthally isotropic distribution.  
We have already seen in Figs.~\ref{fig:fig7} that the 
mean field contribution to the negative $v_2$   
originates mostly from the nucleons of the outer region. This is well quantified
in Fig. \ref{fig:fig16}. The density gradient is higher in the vicinity of
the tips of the overlapping zone of the colliding system. This creates a
stronger force and hence a higher momentum flow.
The out-of-plane flow, created by the mean field, has reached a maximum at half the passing time  
for the reaction at the lower energy, $0.6$ AGeV,  and at passing time for the
higher energy. Later it decreases  
due to the formation of the in-plane ridge seen in Fig. \ref{fig:fig14} and
due to the mean field which lowers the momenta of the escaping nucleons. 
Asymptotically, the potential interactions are the main origin of the overall
out-of-plane elliptic flow, $v_2$,  
apart from reactions at energies below 1~AGeV where the collisions
contribute equally when the nuclear matter EoS is soft, i.e. the number of
collisions is large. 

The present scenario is very different from that at ultra-relativistic
energies where the highly compressed overlap region develops a positive $v_2$
which is scaling with eccentricity of the almond shaped 
overlap region which is converted by the pressure gradient into a momentum asymmetry after the resulting expansion.
At low energies, the internal Fermi motion of the nucleons is of the same order of magnitude
as  the momentum changes due to the density gradients. The passing time is long
and the nucleons in the overlap region react to the sudden increase in density  by expanding 
while projectile and target remnants are passing. The higher the beam energy the shorter is the passing time 
and the less the initial Fermi motion inside the projectile and the target can change the shape of the overlap region -
which becomes therefore almost frozen.  
At lower energies, the initial Fermi motion overwhelms the less energetic fireball 
at the outer part of the high density region, making the final momentum distribution almost spherical, 
whereas the inner core remains almond shaped. This latter is not dense enough to create the
pressure necessary to convert the spatial eccentricity into a positive $v_2$ by the consecutive expansion.
The higher the beam energy the more energy is stored in the
overlap region, hence the higher gets the pressure. As a consequence, with increasing the beam energy, 
$v_2$ becomes positive, as also observed experimentally.  

\begin{figure} 
  \centering
  \includegraphics[width=0.95\columnwidth]{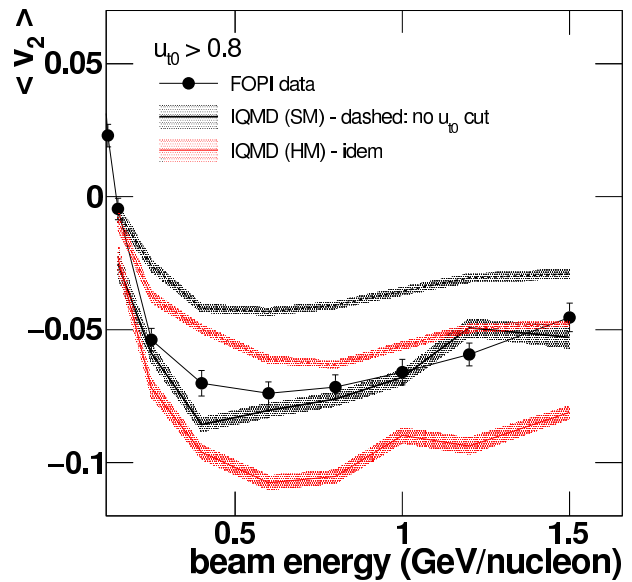}
  \caption{(Color online) Excitation function of the  elliptic flow $v_2$ of
    protons at mid-rapidity. The experimental data (black circles) are from the FOPI
    collaboration published in Fig.~29 of Ref.~\cite{FOPI:2011aa}. The data
    is measured in the impact parameter range $3.1 fm<b<5.6 fm$ and a cut on
    $u_{t0}>0.8$ is applied. IQMD Model results 
    are presented for two different nuclear EOS' (HM with red lines and SM
    with black lines)
    for $b = 4 fm$ and with an additional cut on $u_{t0}>0.8$ (full 
    lines) and without any cut (dashed lines). }
\label{ring_FOPI}
\end{figure}

The excitation function of the elliptic flow parameters $v_2$ of mid-rapidity protons in
\AuAu collisions at b=4 fm is shown in Fig. \ref{ring_FOPI}: The momentum
integrated distribution is shown (dashed lines) as well as the $v_2$ when
requiring that $u_{t0}>0.8$
(full lines). Results with a soft (SM,
black lines) and a hard (HM, red lines) EoS vary widely above 0.4 AGeV beam
energy. We observe in addition a strong beam energy dependence of the elliptic
flow signal in this regime. A maximum of the amplitude is reached
at around 0.6 AGeV. The strength of $v_{2}$ is enhanced when focusing on protons with a large
transverse velocity.  
Comparing with experimental observations for protons  having a high $u_{t0}>0.8$ \cite{FOPI:2011aa} 
at around the same impact parameter, we find a good
agreement using the soft (SM) EoS (full black line in Fig. \ref{ring_FOPI}) 
in accordance with results of Ref.~\cite{Fevre:2015fza}. 
There, both the amplitude and the evolution of the elliptic
flow with the bombarding energy are well reproduced by the model. 

From this analysis we can conclude that the elliptic flow observed in the
reactions around $E_{kin} \approx 1$ AGeV for protons at mid-rapidity
($|y_{0}|<0.2$) has two origins: The collisions of
participant nucleons with the spectator matter 
(collisional contribution) and the acceleration of participants in the mean field (mean field contribution).
The collisional component of $v_2$ is almost independent of
the EoS, whereas the mean field contribution is for a hard EoS (HM) roughly twice as large as
that for a soft EoS (SM). At lower energies (0.6 AGeV) for a soft EoS collisional and 
mean field contributions are about equal, in all other cases the contribution of the mean field
dominates. The mean field induces an out-of-plane flow because those nucleons which are close to
the surface of the interaction zone in y-direction get accelerated out of
the reaction plane due to a strong density gradient in this direction whereas
nucleons close to the surface of the interaction zone in x-direction see a
much smaller density gradient due to the presence of the spectator
matter. This effect is amplified if one selects particles with a high
transverse velocity. 
The calculations with a soft EoS (SM) are in better agreement with the experimental data
than that with a hard equation of state (HM).

\section{Summary}
We analyzed the origin of the experimentally observed  negative elliptic
flow which develops at mid-rapidity in heavy ion reactions 
in the $E_{kin} \approx 1$ AGeV region. QMD calculations have shown that this
elliptic flow depends stronger on the nuclear EoS than any other observable
investigated so far. We have demonstrated that the EoS dependence of this
negative $v_2$  is created by nucleons which are 
situated in the outer part of the overlap region of projectile and
target. Between the maximum overlap and the passing time 
these nucleons experience a weaker
density gradient in the reaction plane as compared to out of the reaction
plane, due to the presence of the spectators.
This translates into a stronger force into the y-direction. The density gradients and
consequently the forces are stronger for a hard EoS (HM) as compared to a soft
one (SM). This explains quantitatively the dependence of $v_2$ on the hadronic
EoS. The scattering of participant protons with the spectator matter produces
an elliptic flow as well, but this component is almost independent of the EoS.
The agreement of the QMD calculations with data for a soft EoS  adds to the
circumstantial evidence that the soft EoS describes correctly the matter at a
density obtained by beam energies of the order of 1 AGeV, an observation that
has already been made by analyzing the $K^+$ production data
\cite{Hartnack:2011cn,Hartnack:2005tr}.

{\bf Acknowledgments:} We acknowledge extensive discussions with
W. Reisdorf. The project was supported by the French-German Collaboration
Agreement IN2P3-DSM/CEA-GSI.

\end{document}